\documentclass[10pt, a4paper, conference, compsocconf]{IEEEtran}

\usepackage[bookmarks=false]{hyperref}
\usepackage{url}
\usepackage{graphics}
\usepackage{epsfig}

\usepackage{xspace}
\let\OldTexttrademark\texttrademark
\renewcommand{\texttrademark}{\OldTexttrademark\xspace}

\begin{document}

\title{SensingKit: Evaluating the Sensor Power Consumption in iOS devices}

\author{\IEEEauthorblockN{Kleomenis Katevas, Hamed Haddadi, Laurissa Tokarchuk}
\IEEEauthorblockA{Queen Mary University of London, UK\\
\{k.katevas, hamed.haddadi, laurissa.tokarchuk\}@qmul.ac.uk}}

\maketitle

\begin{abstract}

Today’s smartphones come equipped with a range of advanced sensors capable of sensing motion, orientation, audio as well as environmental data with high accuracy. With the existence of application distribution channels such as the Apple App Store and the Google Play Store, researchers can distribute applications and collect large scale data in ways that previously were not possible. Motivated by the lack of a universal, multi-platform sensing library, in this work we present the design and implementation of \emph{SensingKit}, an open-source continuous sensing system that supports both iOS and Android mobile devices. One of the unique features of SensingKit is the support of the latest beacon technologies based on Bluetooth Smart (BLE), such as iBeacon\texttrademark and Eddystone\texttrademark. We evaluate and compare the power consumption of each supported sensor individually, using an iPhone 5S device running on iOS 9. We believe that this platform will be beneficial to all researchers and developers who plan to use mobile sensing technology in large-scale experiments.

\end{abstract}


%
\IEEEpeerreviewmaketitle

\section{Introduction}
\label{sec:introduction}

The ubiquity of smartphones as well as the variety of their on-board sensors have enabled the automated acquisition of large scale data, inspiring a wealth of research opportunities. Mobile operating systems such as Android and iOS provide application programming interfaces (APIs) to access these sensors. Lane et al.\ \cite{Lane:2010ek} in a recent survey paper discussed the importance of continuous sensing among different mobile platforms. Various mobile sensing frameworks have been designed that provide continuous sensing, like MobiSens \cite{Wu:2012jr}, EmotionSense \cite{Rachuri:2010bz}, Funf \cite{Aharony:2011fb} and AIRS \cite{Trossen2013}. However, these platforms are currently limited to work on Android or Nokia Maemo phones, limiting the sampling space of users participating in different studies. Since Android and iOS are the two main players in the mobile ecosystem, there is a clear need for supporting continuous sensing in these two mobile environments.

In 2014, we released an early prototype of SensingKit framework \cite{Katevas:2014sk}. SensingKit is a continuous sensing framework compatible with both iOS and Android platforms that enables capturing motion, orientation, location, proximity between devices as well as environmental data from all available sensors inside a smartphone device. Since the two operating systems are equipped with sensor fusion techniques, both raw measurements and fused data like Linear Acceleration and Gravity are supported. Furthermore, SensingKit can also be configured to capture user's natively-labelled activity in supported devices, classified as \emph{Stationary}, \emph{Walking}, \emph{Running}, \emph{Driving} and \emph{Cycling}.

Beside the multi-platform characteristic, SensingKit has some unique features that are not available in other sensing libraries. It fully supports the Bluetooth Low Energy (BLE) specification, branded as Bluetooth Smart (v4.0), for capturing the proximity between devices or other Bluetooth Smart beacons. This has significantly reduced power consumption and highest sampling rate compared to the classic Bluetooth. At this moment, it supports Apple's iBeacons\texttrademark, as well as the new Google Eddystone\texttrademark beacons. These are protocols developed by Apple and Google respectively, that allow a device to broadcast its presence to nearby devices. The receiver can estimate the proximity of the beacon based on the Received Signal Strength Indicator (RSSI) combined with the broadcast \emph{Measured Power} level, the beacon's signal strength measured in 1 meter distance. That feature makes beacon technology extremely useful for indoor localisation systems, allowing smartphones to estimate their approximate location in indoor environments.

In order to avoid timing issues when the user, or even when the device itself changes the system time, the timing in the sensor measurements depends on the device's CPU time base register rather than the system's clock. The library also makes use of the devices motion co-processor for its motion activity recognition sensor, having only a minimum affect on the device's battery life. Finally, it utilises all sensor fusion technology that is available into the operating system, providing calibrated and accurate sensor data.

The framework has already been used in various research projects, including a study that investigates the subconscious phenomenon of gait synchronisation between individuals \cite{Katevas:2015ws}, as well as other Quantified Self applications \cite{Haddadi:2015dqs}. A mobile app titled \emph{CrowdSense} for iOS and Android was also released that facilitates other researches in Mobile Sensing area. By utilising SensingKit, it is capable of collecting sensor data into the device's memory in CSV format.

In the first release, we introduced an early prototype of SensingKit. In that version we only supported Accelerometer, Gyroscope, Magnetometer, Location, Proximity (using iBeacon\texttrademark technology) and Battery sensors. The configuration of these sensors was not possible and the data was in fixed CSV format. Additionally, a universal API between the two platforms was not available, and error handling was limited, making the developing experience difficult.

In this paper, we present the first stable version of SensingKit framework (v0.5) for both iOS and Android platforms. In this new version we have added support for the latest mobile operating systems (Apple iOS 9 and Android Marshmallow) as well as for Apple's new Swift 2 programming language. A universal API now exists and it is fully documented on SensingKit website. We have added extended error handling based on each platforms coding guidelines. Sensors can now be dynamically configured and data can be extracted in both CSV and JSON format. Finally, we have added support for additional sensors including Google's Eddystone\texttrademark, Bluetooth Classic, Screen Status, Pedometer, Altimeter, Microphone as well as other environmental sensors such as Air Pressure and Humidity.

Our objective in this work is to provide an easy-to-use sensing framework that developers and researchers can use to provide continuous sensing in iOS and Android applications. In Section~\ref{sec:arch}, we present the System Architecture and technical details of the framework. The system was evaluated, as reported in Section~\ref{sec:eval}, by measuring the battery consumption of each supported sensor separately, running on an iPhone 5S mobile phone. In Section~\ref{sec:conclusion} we present the conclusions and discuss the future research in this space.

\section{Platform Architecture}
\label{sec:arch}

SensingKit is a modular mobile framework developed in the native programming language of each platform (Java for the Android and Objective-C for the iOS version). It supports mobile devices running iOS 8 and Android Jelly Bean (v4.1) and above. At this moment, that corresponds to 95\% of all iOS and 95\% of all Android devices available today\footnote{As reported by Apple App Store and Google Play Store on March 7, 2016}.

\begin{figure}[!t]
    \centering
    \epsfig{file=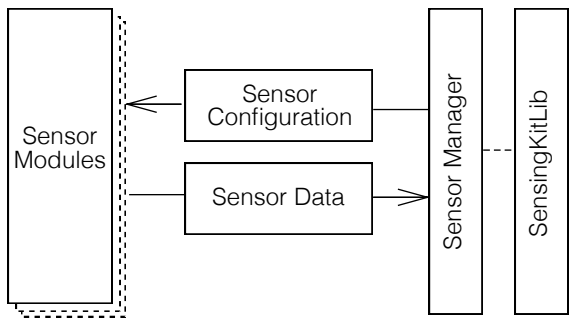}
    \caption{SensingKit System Architecture}
    \label{fig:architecture}
\end{figure}

Figure~\ref{fig:architecture} gives an overview of the system architecture. For every sensing category, a sensing module (e.g. SKAccelerometer) exists in SensingKit, as well as a corresponded configuration (e.g. SKAccelerometerConfiguration) and a data object (e.g. SKAccelerometerData). Each sensor module provides access to the corresponding sensor inside the device whereas a configuration object initialises the sensor with custom configuration (e.g. custom sample rate or accuracy). When new sensor data is available, a data object is generated that represents the sensor data in CSV or JSON format. SensingKitLib is the interface that developers need to use in order to check for sensor availability inside the device, initialise and configure a sensor, provide the block function that will be called each time new sensor data is available, and finally start or stop continuous sensing operations. Sensor Manager is the module that implements SensingKitLib interface and performs all required operations to the sensor modules such as memory allocation and deallocation, sensor configuration etc. It is important to mention that due to the modular design of this library, it is easy to develop a new module and extend its sensing capabilities. Table~\ref{tab:supported_sensors} presents the available sensing modules of the framework. 

\begin{table}[!t]
\centering
\caption{SensingKit: Supported Sensors}
\label{tab:supported_sensors}
\begin{tabular}{ l c c }
    \hline
    Sensor               & Apple iOS              &  Google Android  \\
    \hline
    Accelerometer        & Yes                    &  Yes  \\
    Gravity              & Yes\textsuperscript{*} &  Yes  \\
    Linear acceleration  & Yes\textsuperscript{*} &  Yes  \\
    Gyroscope            & Yes                    &  Yes  \\
    Rotation             & Yes\textsuperscript{*} &  Yes  \\
    Magnetometer         & Yes                    &  Yes  \\
    Pedometer            & Yes                    &  Yes  \\
    Altimeter            & Yes                    &  Yes  \\
    Humidity             & -\textsuperscript{**}  &  Yes  \\
    Light                & -\textsuperscript{**}  &  Yes  \\
    Ambient Temperature  & -\textsuperscript{**}  &  Yes  \\
    Location             & Yes                    &  Yes  \\
    Motion Activity      & Yes                    &  Yes  \\
    Battery              & Yes                    &  Yes  \\
    Screen Status        & Yes                    &  Yes  \\
    Microphone           & Yes                    &  Yes  \\
    Bluetooth\textsuperscript{\textregistered} Classic & - & Scanning only \\
    iBeacon\texttrademark Proximity        & Yes           & Yes  \\
    Eddystone\texttrademark Proximity      & Scanning only\textsuperscript{***}  & Yes  \\
    \hline
    \multicolumn{3}{p{8cm}}{\textsuperscript{*}\footnotesize{In SensingKit-iOS, these sensors are part of the \emph{Device Motion} sensor.}} \\
    \multicolumn{3}{p{8cm}}{\textsuperscript{**}\footnotesize{These sensors are either not available or access is not allowed on iOS due to Apple's restrictions.}} \\
    \multicolumn{3}{p{8cm}}{\textsuperscript{***}\footnotesize{Broadcasting an Eddystone\texttrademark beacon signal is not allowed on iOS due to Apple's restrictions.}}
\end{tabular}
\end{table}

For proximity sensing, SensingKit uses the new Bluetooth Smart (4.0) proximity profile. This profile allows to broadcast a device's presence, scan for other devices and most important, estimate the distance between them using the Received Signal Strength Indicator (RSSI). Bluetooth Smart is only fully supported in Android Lollipop (v5.0) mobile operating system. Android Jelly Bean (v4.3) devices are only limited to scan and connect to other devices (\textit{Observer and Central mode}) and not to advertise its presence to the nearby devices (\textit{Peripheral mode}). Due to Apple restrictions applied to iOS operating system, it is not possible to broadcast a Bluetooth Smart signal with custom service data blocks, making it impossible to broadcast an Eddystone\texttrademark signal from an iOS device.

\section{Evaluating the Battery Consumption}
\label{sec:eval}

\begin{figure*}[!ht]
\centering
\includegraphics[width=7in]{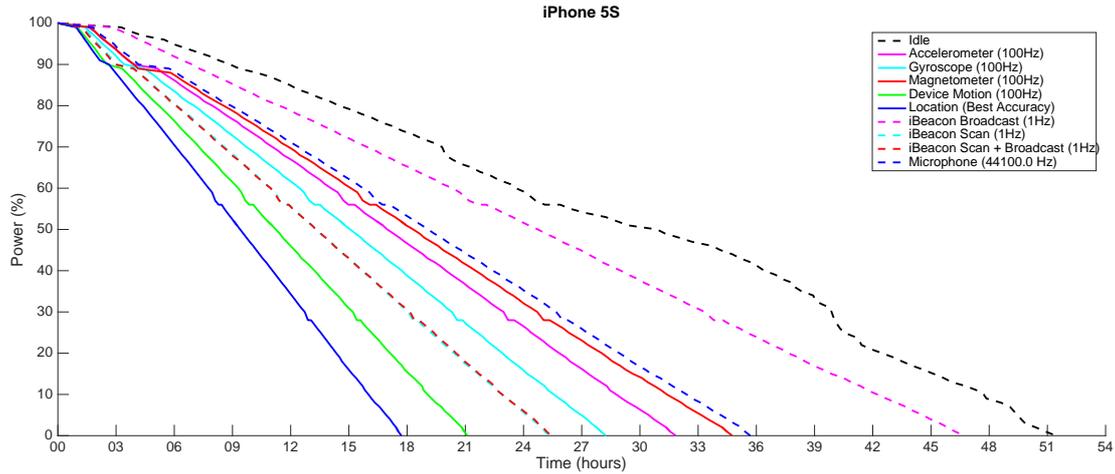}
\caption{Battery consumption of SensingKit running on an iPhone 5S.}
\label{fig:battery}
\end{figure*}

We measured the battery life performance while using SensingKit in an iPhone 5S device running iOS 9.0.2 (Table~\ref{tab:specification}). The device was fully erased and restored into the manufacturers default settings. No other third-party applications ware installed or running in the background. The device was set to Flight Mode, having Wi-Fi, Bluetooth and Cellular connectivity disabled. Finally, the Background App Refresh setting was set to Off and the Low Power Mode to On, in an attempt to minimise the impact that the operating system has on the device's battery life.

\begin{table}[!ht]
\centering
\caption{Device Specification}
\label{tab:specification}
\begin{tabular}{ l | l }
    Model            & iPhone 5S  \\
    Storage          & 32 GB      \\
    Operating System & iOS 9.0.2  \\
    Processor        & 1.3 Ghz Dual-core \\
    Memory           & 1GB LPDDR3 \\
    Battery          & 1560 mAh   \\
    Bluetooth        & 4.0        \\
\end{tabular}
\end{table}

Figure~\ref{fig:battery} and Table~\ref{tab:consumption} show the energy consumption of SensingKit running on the mobile device described above. We show the consumption of the library while using the Accelerometer, Gyroscope, Magnetometer, Device Motion (fused motion and orientation data), Location (GPS), iBeacon\texttrademark and Microphone sensors. In the case of iBeacon\texttrademark sensor, we first evaluate Broadcasting and Scanning modes separately, and then a combination of both of them together. In addition, we visualise the library running in ``idle'' mode, when it only senses the battery levels.

\begin{table}[!ht]
\centering
\caption{Battery Consumption using SensingKit for iOS}
\label{tab:consumption}
\begin{tabular}{ l c c }
    \hline
    Sensor                     & Sample Rate   & Hours Lasted \\
    \hline
    Idle                       & -             &  51.27    \\
    Accelerometer              & 100 Hz        &  31.51    \\
    Gyroscope                  & 100 Hz        &  28.15    \\
    Magnetometer               & 100 Hz        &  34.45    \\
    Device Motion              & 100 Hz        &  21.07    \\
    Location                   & Best Accuracy &  17.42    \\
    iBeacon Broadcast          & 1 Hz          &  46.43    \\
    iBeacon Scan               & 1 Hz          &  25.21    \\
    iBeacon Scan \& Broadcast  & 1 Hz          &  25.26    \\
    Microphone                 & 44100.0 Hz    &  35.41    \\
    \hline
\end{tabular}
\end{table}

The results show that the Location (GPS) sensor in ``Best Accuracy'' mode is the most power expensive sensor of all, as the device only lasted for 17.42 hours compared to the ``idle'' mode that lasted for 51.27 hours. GPS sensor is well known for its extensive power consumption, not only because it receives signal from multiple satellites simultaneously in order to estimate the devices distance from them, but also because of the expensive trigonometric operations (trilateration) that is performing in order to estimate the device's position on the surface of the earth.

From all motion and orientation sensors, Magnetometer is the one that performed best, as the device lasted for 34.45 hours in 100 Hz sampling rate. Accelerometer came next, sensing motion data in 100 Hz and lasting for 31.51 hours, where as Gyroscope lasted for 28.15 hours in the same sampling rate. As expected, the Device Motion sensor is the most expensive of all motion sensors, lasting for 21.07 hours. The reason is that this sensor is using a combination of Accelerometer, Gyroscope and Magnetometer in order to provide calibrated and more accurate data using sensor fusion techniques performed entirely on hardware.

Recording audio using the Microphone sensor lasted for 35.41 hours, despite its high sampling rate of 44100.0 Hz.

Evaluating the iBeacon\texttrademark sensor in the three different modes explained above showed interesting results. While the sensor was set in the ``broadcast'' mode, the device lasted 46.43 hours, highly comparable to the ``idle'' mode (51 hours). More interestingly, there were only 5 minutes difference between the ``scan'' and ``scan and broadcast'' modes, as the devices lasted 25.21 and 25.26 hours respectively. That proves that broadcasting an iBeacon\texttrademark signal has almost no effect on the device's battery, while scanning for other iBeacon\texttrademark devices is quite expensive. The reason is that iOS not only scans for the presence of other devices, but is also ``ranging'' in 1 Hz sampling rate in order to estimate the other beacon's proximity based on the RSSI explained above.

It is important to mention that Figure~\ref{fig:battery} only represents the battery consumption on the specific mobile device listed in Table~\ref{tab:specification} and should only be viewed as a comparison between the available sensors rather than an indicator of each sensor's power consumption.

\section{Conclusions and future work}
\label{sec:conclusion}

In this work, we have presented an extension on SensingKit, a continuous sensing system that works in both Android and iOS environments. We evaluated the battery consumption while using Accelerometer, Gyroscope, Magnetometer, Device Motion, Location, iBeacon\texttrademark and Microphone sensors on an iPhone 5S smartphone. We plan to continue the development of this framework and extend its sensing capabilities. More specifically we plan to adopt mobile health sensors by supporting HealthKit on iOS and GoogleFit on Android. We believe that this work will be beneficial for researchers willing to conduct large-scale experiments using mobile sensing.

More information about SensingKit as well as the complete source-code is available at \href{http://www.sensingkit.org}{www.sensingkit.org}.

\section*{Acknowledgement}

This work is supported by funding from the UK Defence Science and Technology Laboratory.


\newpage

\bibliographystyle{IEEEtran}
\bibliography{IEEEabrv}

\end{document}